# Security in Monitoring Schemes: A Survey

Atul Vaibhav
Student
TU Darmstadt
Darmstadt, Germany
`Atul.vaibhav@stud.tu-darmstadt.de`

## ABSTRACT
With our growing reliability on distributed networks, the security aspect of such networks becomes of prime importance. In large scale distributed networks it becomes cardinal to have an efficient and effective monitoring scheme. The monitoring schemes supervise the node behaviour in the network and look out for any discrepancy. Monitoring schemes comprise of monitoring components that work together to help schemes in meeting various security requirement parameters for the networks. These security parameters are breached via various attacks by manipulation of monitoring components of particular monitoring schemes to produce faulty results and thereby reducing efficiency of networks, reliability and security. In this paper we have discussed these components of monitoring, multiple monitoring schemes, their security parameters and various types of attacks possible on these monitoring components by manipulating assumptions of monitoring schemes.

## 1. INTRODUCTION
In the journey from centralized networks to distributed networks, there have been significant changes in people's life and a growing dependency on distributed networks. These networks provide means of reliable, robust and scalable communication. With the increasing popularity, the dependence on these networks is growing day by day, and hence the concern for security of these networks is coming in lime light very fast. The implementation of distributed communication networks in various critical domains of our society has urged a lot of research in direction to improve the security aspects in monitoring against attacks.

There are multiple implementations of distributed networks. Two of the most popular implementations of it are – Mobile Ad Hoc Networks (MANETs) and Wireless Sensor Networks (WSN). The wireless networks formed by self autonomous collection of mobile users that communicate over relatively bandwidth constrained wireless links are called mobile ad hoc network. The services provided in this form of networking ranges from transfer of data for chat, videos to data storage etc. and because of the ease of use and deployment they have implementations in establishing survivable, efficient, dynamic communication for emergency/rescue operations, disaster relief efforts, and military networks also.

Wireless Sensor Network (WSN) is a group of small, light weight, low cost sensor network, with small power intake connected to each other to accumulate data relating to an event occurrence like movement, change in temperature, pressure, and multiple such physical changes. These sensors are deployed in large numbers to form an ad hoc network that work in a particular direction for a single goal of monitoring and tracking changes in their environment with cooperation of other sensors. Despite being of low computational capabilities, less resources, small memories and limited network capabilities, they work together as a unit and find multiple applications in field of road traffic monitoring, military domain, medical, environmental monitoring, surveillance in battlefield, robotics and multiple other critical areas.

The network topology in distributed networks is highly dynamic, and hence they need to be continuously monitored to update network structure and predict any type of malicious activity among the network nodes. There multiple monitoring schemes that have been discussed in past for this. These monitoring schemes follow pattern of data collection, interpretation of data and dissemination of data in the network. These processes are vulnerable to multiple types of attacks on all three levels. These attacks range from stale data, manipulation of analysis of data to malicious data circulation.

For example there is a Fire Detection service implemented in WSN in an area as briefly discussed in [8][24]. The multiples nodes in that area report their changes in temperature to the superior node, or the querier, depending upon the monitoring schemes implemented. In tree based structure the nodes a particular zone report to the immediate manager node about the events and group of managers report to the senior node and this continues till the information reaches the base station which in turn locates the area and informs the Fire station the particular location. In case of fire the nodes notice the increment in the temperature and report to the manager, manager then identifies the area affected with fire from the location of nodes reporting high change in temperature in their zone and report to higher manager which further locates the zones affected with fire. The base station on receiving information processes and analyses the response for area, criticality of zones affected and removes malicious or fake information and disseminates data back to network for changes in topology, new routing information etc. and eventually reports to external user that is in this case the fire station. This way the information is collected, processed and disseminated in a Network.

In this paper we have tried to briefly discuss monitoring schemes, the structure of the network information flow topology, processing for monitoring and classification to some of the most popular attacks in this process. The study of these attacks provides us a broader view for the attack handling and providing resilient monitoring schemes for a secure distributed communication system.

## 2. DISTRIBUTED MONITORING

### 2.1 Distributed Monitoring Components

Monitoring process is carried out to collect the data from the nodes to infer the results and detect any malicious activity or any type of attack being carried out in the network. This data is finely processed and the results are sent out to the nodes regarding the next course of action, such as removal of malicious nodes from the network, adding resources, routing information etc.

Typically monitoring process has three components:

**a) Aggregation**

As per [9] Data aggregations work on a simple source and sink methodology. Nodes in distributed networks are generally have uneven resource distribution and with limited capacity for storage and computation. To avoid large amount of data transmission over network, the data from nodes is aggregated at nodes before sending it further in smaller sizes to reduce energy and resource consumption of nodes in network. The nodes in the distributed network act as sources while the base station or the querier who initiates the query, collects the data from the sources and act as sink. The aggregation starts at the base station, where it generates query for data aggregation from the nodes and broadcasts it on complete network or subset of a network. Now the source nodes collect the information to be sent for aggregation, and send it to the aggregator broadcasted. Now this aggregator receives the data, now instead of sending all the data to the base station, it aggregates the data into small packet on basis of an aggregation function and transmits it to the base station for the calculation and interprets meaningful result out of the data. Now the nodes in network collects the information relating changes in their environment and network structure based on required aggregation function and sends this information to upper querier and so on back to the base station who initiates the query for further processing.

**b) Decision Making/Analyzing (Based on collected data)**

The data collected by the querier from the source nodes is processed by the base station or in some cases any external user who has access to the network. The data is processed with limited computational capabilities to fathom the events occurred in the addressed field. It is hard to process and send large amount of data on the communication lines, as it generates large overhead for nodes with limited power and resources. To overcome this aggregation functions are used to reduce energy consumption by reducing the packet size for transmission of data and easier processing. The data processing is actually implementation of the aggregation function and implementation of decision making process [15, 16]. In aggregation function we collect the values from the source nodes and process them before being forwarded to the other node by implementation of small level functions like mode, mean, median, average etc. These small functions are implied on the collected data from other nodes and data from the processing node itself, to generate some generalized value for all data. Now this data being of the similar size it sent easily over the network and is easily computable by receiver node.

**c) Dissemination**

Very often it is needed to update the information on the nodes in the network or update functionalities of the nodes. This needs large amount of data dissemination over the network to every node. As discussed in [6] Dissemination of data is required for proper result in network for data aggregation and to update the stale routing information as the network is highly dynamic in nature and multiple nodes leave, stop working or declared malicious or join the network. Dissemination also serves very important purpose of time synchronization and multicasting in the network. Based on the data aggregated and decision making/ analysis, all the nodes need to be updated regarding multiple parameters regarding calibration of sensors, updating network parameters etc [17]. Energy consumption is a major issue for disseminating large data objects over network and multiple protocols have been developed to achieve higher efficiency and security in it [1].

### 2.2 Classification of Monitoring Schemes

There have been multiple monitoring structures defined previously to implement the monitoring process. These structures provide the topology of information flow in the distributed networks. There are basically three most popular structures in monitoring schemes:

#### 2.2.1 Tree Based Monitoring Scheme

Tree based monitoring scheme is also known as the Hierarchal system. According to scheme discussed in [9] each geographical area is divided into smaller areas or zones; each zone has hierarchy of managers. Each domain manager responds only to its superior manager. In case of any event or information generation, lower nodes send it to the higher nodes acting as managers for information aggregation, till the top of the chain, while information dissemination occurs in opposite manner. Nodes continuously communicate with each other in a timely fashion. Time interval for responding is periodic. After every certain time interval periodic responding time is reduced by one. If there is problem in response from lower level then the area manager spawns a new node that replaces old one and informs related nodes. This is replacement method. But in case when manager fails one of the lower levels is promoted to the manager level. This method is most efficient in information flow but lacks robustness.

#### 2.2.2 Gossip Based Monitoring Scheme

As per discussed in [13] Gossip based monitoring scheme is analogous to our normal gossiping activities, where one person tells other person about news and the news spreads in a network. Similarly the nodes in distributed network interact with each other spreading the information to the local neighbors. Now these neighbors send information to randomly select other neighboring nodes until the information is spread in the entire network. This scheme includes periodic or event based interaction among nodes where each nodes shares information in bounded size interactions. Along with the gossip they also share liveliness information on regular basis. When one node fails to receive the liveliness information of another node directly from that node directly or as gossip from the other nodes in the network then the node is suspected to have failed.[17] When all the nodes in the network suspect failure of that node then that node is declared dead in the network.

*2.2.3 Hybrid Monitoring Scheme*

Hybrid schemes follow the pattern of both tree based and gossip based schemes. In hybrid scheme there is integration of both tree and gossip based topology of information flow to utilize the advantages of both the schemes. S. Idreos proposed the structure of a hybrid monitoring scheme [10] as a complex mix of both the topologies where a group of nodes following gossip based monitoring communicate with each other and help in decision making process. This group of nodes act as Super nodes and rest of the nodes connected to them are Client peers, and all client peers interact in tree based manner. Hybrid schemes may implement different topology for different monitoring components, e.g. Nodes may disseminate data in hierarchal manner and rest of communication and processing in gossip, thereby utilizing the both of best world [16].

# 3. SECURITY REQUIREMENTS

- **Data Integrity**: is required to make sure the data is not changed or altered in any way. Any manipulation of data could occur because of external attack, transmission failure or by accidental damage to data. The attacker can alter the data sent on a network by adding false content in data package or changing data packets in any way this produces false data while data analysis or modify data to be aggregated and result in wrong data dissemination to nodes in network. Data Integrity is needed to be maintained to avoid any manipulation of data to prohibit its adverse affects on data analysis [8].

- **Data Freshness**: is required to check attacks like replay attack, where stale data is fed to the network. During data aggregation it is important to collect fresh data from the source node to avoid any discrepancy in the aggregation result. And the network should be updated with fresh data, including routing information, updated network topology etc. to synchronize the network during data dissemination [3][8][9]. During data aggregation partially fresh data can be used where the delay in message delivery is not counted but during dissemination the data needs to completely fresh for synchronization.

- **Consistency:** all the nodes need to be on the same page with the network structure and routing updates, but there is a lot of delay, damaged packet and packet dropping which makes it harder to maintain this consistency in distributed networks. This is crucial information that keeps the entire network updated, so it has to be delivered without any data loss or damage to data packet. To make sure every node receives this update, there are no time bound placed on it, hence eventually all nodes receives every byte of the update. In case of any problem in consistency the network might lose the packet without updated routing information and network structure [1][8].

- **Confidentiality:** refers to transmission of data in a way that the unauthorized personnel cannot read the data packet. This is ensured by encryption of data packet sent on the network [9]. Nodes in a network should not leak sensor readings to its neighbors in highly sensitive matters like key distribution hence confidentiality is really important. Public encryption keys should also be encrypted to some extent to protect against traffic analysis attacks. Nodes follow two approaches for encryption of the data packet. In first approach, every node encrypts the data and sends to next node, than the next node receives the data, decrypts the data, applies the aggregation function and again encrypts the data before sending it further to next node. This approach brings a lot of overhead for network and for computational capacity of the nodes. To save this an approach has been defined, with use of homomorphic encryption [11] where the nodes use the encrypted data and apply aggregation function without decryption with application of homomorphic encryption. This way is more suitable for networks with nodes of low computational capabilities and uses less power.

- **Data Availability:** It is of prime importance to make data available all the time in a network. In case any node is compromised or maliciously attacked, it tries to block the availability of data in that area of network. This deprives other nodes of data freshness and halts data aggregation, analysis and dissemination process in network. To avoid this network needs to heal itself to resume the data availability in the network. For the network to survive or be alive the needs to be regularly made available and thus the nodes try to reconnect to heal network and try to pass large amount of data in case of attack [8][9]. Sometime in case of attack, nodes rotate the positions to transmit large amount of data to help network survive.

- **Non-repudiation:** in distributed networks, where there is a lot of flooding of messages, it is important to make sure who is sending and receiving messages, hence non repudiation is very important factor of distributed communication. The nodes sending or receiving the messages must commit to data they are sending or receiving and should not be able to deny [9]. These way nodes could track any false or misleading dissemination of data and during process of aggregation nodes could track the source of changes and deviation for analysis result.

- **Authenticity:** is required in a distributed network to authenticate the any malicious node feeding incorrect data to

other nodes. It is of two types: entity authentication and data authentication [9]. Entity authentication is required in a network to verify the sender, that the data is actually sent by that sender and it is legitimate member of the network and is in possession of the cryptographic key. The nodes also need to authenticate the data packet to verify that the data packet sent by the node is not altered and is original. Both of the authentications are required in network to verify secure data transfer among nodes. In case any node sends fake data by posing as any other node, then entity and data authentication are used to verify the legitimacy of the information.

- **Secure Localization:** mostly attacks are configured on the basis of location of the attacking node or centre of the malicious activities to narrow down the group of nodes that are responsible. But any attacker node can easily manipulate its location and send to network, thus making it very difficult to locate attacking node. Multiple schemes have been discussed to verify location of nodes like SPINE (Secure Positioning for sensor NEtworks) algorithm, SeRLoc (Secure Range-Independent Localization), VM (Verifiable Multilateration) etc. In verifiable multilateration (VM) location of a node is calculated by a number of verifiable references in the network [8][10]. This scheme locates nodes by bounding the distance of nodes from each other and authenticates the ranges. In case any attacking node sends false information about its location then it would need to change the distance from reference node, but then it would need to also alter the distance from other nodes because of authentication ranging. This way it is easier to location any activity in network and making it secure.

## 4. ATTACKS ON MONITORING COMPONENTS

This section includes multiple types of attacks and how they manipulate the monitoring components to attack the monitoring schemes. Generally attacks manipulate aggregation, analysis and dissemination components to generate or send wrong, stale and manipulated information to affect the monitoring process. Some of the most popular attacks in this context are:

1. **Faulty Reading attack:** as discussed by [19] this attack focuses on the faulty readings from other nodes in the network. These faulty readings arise from arbitrary readings sent by faulty nodes or noisy readings arising from interferences [20] these readings hamper the very purpose of monitoring schemes. To counter this most of the monitoring schemes filter out these faulty readings based on the statistical analysis of the data received from various nodes and filter out the unusual ones. But this approach has a weakness. It may compromise the accuracy of the monitoring scheme by accidently filtering out some important event.

   Another method used by monitoring schemes as mentioned in [21] is spatial correlation. It suggests that in case a node receives any faulty reading, it uses spatial correlation and finds the neighboring nodes of the sender of faulty reading. It sends the faulty reading to these nodes which then compare it to their reading and if the difference is higher than a predefined limit, they send back a negative vote and if it is below it sends back a positive vote. If number of votes is higher for negative then the reading is discarded and if positive is higher than this value is not discarded. Despite effectiveness of this scheme there are few weakness which can be used by attacker. First, [19] spatial correlation focuses on the distance between nodes as a reason for giving similar values or readings while distance cannot fully ascertain relation between values of two nodes and hence may have a large value variance. Secondly, in case nearest nodes are compromised then the result of voting may also be faulty, as the trust value of these nodes are not tested.

2. **Accuracy Attack:** In networks congested networks it is very problematic to replay data securely to neighbors and accuracy problem arises. In hierarchical topology the data is collected by cluster heads. The nodes from which data gets delayed, due to low data rate creates accuracy problems for data aggregation. The measured values reach the cluster heads in untimely fashion causing a difference in aggregated value. Now high variance in data rate causes nodes to result with possible stale information and disrupts the updation of node value. This problem degrades the efficiency of data aggregation. Solutions of this problem have been discussed thoroughly and many solutions have been provided like SLB Protocol [23] and others.

3. **Denial of Service Attacks (DoS):** one of the most popular attacks performed by jamming the network. As in [7], it is carried out by putting large number of data packet on network or modifying the routing structure to channel huge number of data packet from the same communication line, or by interfering the network line by using same radio frequency that the network channel is using to disrupt the packet flow in that frequency. This results in network blocking or jamming, which causes refusal of service to the authenticated user because of large number of packet on the network. To aggregate the data packets aggregator nodes try to send data upstream to the receiver node, but because of the service denial, sending fails or complete data doesn't arrive at the receiver node. This fails the whole concept of data aggregation.

4. **Bad Mouthing or Reputation Trap attack:** discussed in [22] distributed Systems today are more secure and use trust and reputation based system in networks to avoid any attack on monitoring schemes. Monitoring schemes use these systems to identify compromised node and faulty data from nodes with low trust and reputation value. Bad mouthing attacks only occur in systems that consider feedbacks and reputation matrices for trust evaluation of nodes. In this attack the attacker node manipulates weaknesses of data analysis and dissemination to improve trust of network in it and decreases trust in other nodes. During dissemination the attacker node manipulates the feedback by sending negative feedbacks about other nodes and positive feedbacks about itself. The systems lacking data authentication, accountability in dissemination of data fall prey to these fabricated feedbacks and alter the reputation of nodes based on fake data. Such attacks are more effectively carried out in groups as it is easier to verify feedbacks in case of integrity and data authentication checks during dissemination. Multiple attacks like *Sibyl attack* or *colluding attack* use this weakness in monitoring system and produce severe ramification in network like Dos. The attackers in colluding attack can further manipulate data aggregation to improve trust value of other nodes. Such attacks can be avoided by collection proper proof for valid transaction and method to avoid Sybil and collusion attacks.

5. **Node Compromise:** it occurs when in a network an external agent take control of the node after its deployment. [2] The attacker can now use this node for multiple purposes to make colossal damage in the network. Generally the attacker finds a node in the network and gain control of its interface and connects to it via its computer or makes connection of some sort. Now the attacker can send fake data, multiple copies of data to the aggregator. The compromised node may even decide to discard or alter the data packet while sending to an aggregator. Compromised node may misdirect the aggregated information and deprive network of important information.

6. **Sybil Attack:** more popularly known as the identity disorder among nodes is an attack which is carried out by a node which acts as a group of nodes by impersonating other nodes in the network or fabricating new ones. The attacker node creates multiple identity of a node to appear as a group of distinct multiple nodes to the network [3]. Because of node redundancy, Sybil attacker node advertises easily its ability to share resources, quality of its route and hence gains trust, and reliability in the network. Once the Sybil node becomes a part of the network it attracts large traffic through it and gains information about a lot of nodes. Now a Sybil node can affect aggregation process severely, as it impersonates a lot of nodes, it can feed a large number of false entries to the aggregator if it manages to generate large number of replies. In case of validation of the entry of data to aggregator, Sybil node may generate large number of responses for validation of entry from its fake identities.

Sybil node can gain trust so it can choose any malicious data aggregator which might impact aggregation process very badly.

7. **Sinkhole Attacks:** In distributed network there are data sinks that collect data from sources; in case of sinkhole attacks, a compromised node act as sink for the data from the sources. This causes that data leakage and packet loss in the network. [12][14] Sinkhole nodes attract the nodes to transfer the data through it, making it look as high promised path with reliability and latency information and thus influence multiple nodes in the network, sometimes by help of Sybil nodes. Now while monitoring this node act as aggregator to receive data from source nodes. These nodes start sending data packets through it and sinkhole manages to get all of the traffic in the targeted area to pass through it. Instead of forwarding these packets sinkhole drops the packets or selectively forwards or modifies the data packets and thereby crippling the aggregation process from a zone completely. Multiple protocols suspect such behavior and monitor such node but still most of the protocols fall prey to this technique.

8. **Replay Attack:** this attack is also known as 'Man in the middle attack'. In this attack the attacker stores some information from the network without any understanding of it; attacker externally attacks the analysis process by replaying this message at a later stage in time. It sends this data on a later stage to the aggregator and manages to change the output. This attack works similar to data entry from the source node to the aggregator, which is stored by attacker node by overhearing or eavesdropping in the network. This entry is replayed by the attacker at any later stage to change the aggregation result and thus changes the result of aggregation [3][8][15].

9. **Selective Forwarding Attack:** or commonly known as the grayhole attack works as the name suggests, the malicious node receive data packet from the previous node but forwards selectively to the next node. While the whole distributed network works on trust and assumption that every node abides by the protocols, such nodes may choose which packet to drop and forward. In response to aggregator's query such node chooses to forward data from a particular node or drop it, thereby generating wrong aggregation analysis by blocking a particular zone or area of networks from sending their data for aggregation [9]. It may act as complete blackhole node where all the received packets are dropped by the nodes but in such case neighboring nodes consider that node is dead and exclude it from the network. To avoid this in such attacks, nodes use a subtle method of selectively forwarding data creating disorder in aggregation instead of completely dropping all the data packets. The selective forwarding results in tempered aggregation value [12][9].

## 5. CONCLUSION

Distributed network is a promising field with unlimited possibility for implementation in our day today life. There has been continuous research going on in developing secure methods of communication and there is still large scope for research in area of security in distributed networks.

The paper provided a general understanding on concept of monitoring in distributed networks and its security. Distributed networks were explained with help of examples. We discussed the components of monitoring and classified monitoring schemes. Further on we explained the security parameters that are needed to address the issue of security in monitoring schemes. Addressing the attacks on monitoring schemes we tried to put light on loopholes manipulated by attackers to affect monitoring components and eventually whole network.

Because of the wide varieties of attacks on multiple levels to manipulate monitoring we found there are a large number of attacks still undisscussed and found a future direction for extending research to address various aspects of security in this filed.